\documentclass[journal]{ieeetran}

\usepackage[dvipsnames,svgnames]{xcolor}

\newcommand{\rOne}[1]{#1}
\newcommand{\rTwo}[1]{#1}
\newcommand{\otherChange}[1]{{#1}}

\AtBeginDocument{%
  \providecommand\BibTeX{{%
    Bib\TeX}}}

\usepackage[english]{babel}

\usepackage{amsmath}
\usepackage{graphicx}
\usepackage{booktabs}
\usepackage{xspace}
\usepackage{xcolor}
\usepackage{comment}
\usepackage{xurl}
\usepackage{multirow}
\usepackage{multicol}
\usepackage{caption}

\newcommand{\chatbot}{CA\xspace}
\newcommand{\chatbots}{CAs\xspace}

\newcommand{\Chatbots}{CAs\xspace}

\newcommand{\quoteP}[1]{``\emph{#1}''}

\def\BibTeX{{\rm B\kern-.05em{\sc i\kern-.025em b}\kern-.08em
    T\kern-.1667em\lower.7ex\hbox{E}\kern-.125emX}}
\begin{document}

\title{Prevalence of Security and Privacy Risk-Inducing Usage of AI-based Conversational Agents}

\author{\IEEEauthorblockN{Kathrin Grosse},
\IEEEauthorblockA{Independent Researcher, Italy} \\
\and
\IEEEauthorblockN{Nico Ebert},
\IEEEauthorblockA{ZHAW,
Winterthur, Switzerland \\
ebet@zhaw.ch}
}
\maketitle

\begin{abstract}
    Recent improvement gains in large language models (LLMs) have lead to everyday usage of AI-based Conversational Agents (CAs). At the same time, LLMs are vulnerable to an array of threats, including jailbreaks and, for example, causing remote code execution when fed specific inputs. As a result, users may unintentionally introduce risks, for example, by uploading malicious files or disclosing sensitive information. However, the extent to which such user behaviors occur and thus potentially facilitate exploits remains largely unclear. To shed light on this issue, we surveyed a representative sample of 3,270 UK adults in 2024 using Prolific. A third of these use CA services such as ChatGPT or Gemini at least once a week. Of these ``regular users'', up to a third exhibited behaviors that may enable attacks, and a fourth have tried jailbreaking (often out of understandable reasons such as curiosity, fun or information seeking). Half state that they sanitize data and most participants report not sharing sensitive data. However, few share very sensitive data such as passwords. The majority are unaware that their data can be used to train models and that they can opt-out. \rTwo{While potentially risk-inducing behavior does not directly translate into actual risk, our findings suggest that current academic threat models might} manifest in the wild, and mitigations or guidelines for the secure usage of CAs should be developed. In areas critical to security and privacy, CAs must be equipped with effective AI guardrails to prevent, for example, revealing sensitive information to curious employees. Vendors need to increase efforts to prevent the entry of sensitive data, and to create transparency with regard to data usage policies and settings.
\end{abstract}

\section{Introduction}
The latest AI-based conversational agents (CAs) make it easy for end users to interact: many CAs take natural language, images and other files as input and respond in natural language, generated images, program code, and other outputs. For example, ChatGPT, which launched in November 2022, gained more than 100 million users in its first two months - the fastest growing user base of a consumer application to date.\footnote{\url{https://www.reuters.com/technology/chatgpt-sets-record-fastest-growing-user-base-analyst-note-2023-02-01/}} According to the Pew Research Center, as of February 2024, 23\% of U.S. adults had already used ChatGPT~\cite{mcclain_americans_2024}. Open AI claims that by August 2023, 80\% of Fortune 500 companies were using its product~\cite{chatgpt2023}.

However, large language models (LLMs) and large multimodal models (LMMs) - the foundation of CAs - are vulnerable to attacks. In security~\cite{yao2024survey, abdelnabi2023not, perez2022ignore, bagdasaryan2023ab} and privacy~\cite{yao2024survey, pan2020privacy, li2023multi}, these models are often described as a double-edged sword. CAs can therefore support tasks such as strengthening software code security or safeguarding data confidentiality~\cite{yao2024survey}, but also inherit the broader vulnerabilities of AI systems~\cite{biggio2018wild, papernot2018sok}.

Many security and privacy risks can be unknowingly created by ordinary users, both at home and at work. For example, text inputs~\cite{perez2022ignore} or image uploads~\cite{bagdasaryan2023ab,fu2023misusing} may lead to vulnerabilities, with unexpected side effects on the underlying LLM or LMM. The integration of CAs with other software, for example, in operating systems~\cite{apple} or in private and business contexts may lead to ``spillover effects'' and aggravate the harm of malicious inputs~\cite{abdelnabi2023not}. Users may also try to intentionally ``jailbreak'' \chatbots, making the model output unintended information~\cite{wei2023jailbroken,liu2023jailbreaking}. End users might share personal or other sensitive data with \chatbots~\cite{forbes}, too. The human-like interface of CAs might even amplify data sharing ("anthropomorphizing")~\cite{zhang2023s}. 

Even if it is not yet clear how likely a behavior is to pose a risk, it is necessary to examine the prevalence of potentially risky behaviors in order to assess risks~\cite{ACSFBS24}, support the development of mitigation measures such as AI guardrails~\cite{dong2024position} and establish effective control mechanisms for users and organizations.

\textbf{Contributions.}
The goal of our explorative survey was to create a first, descriptive snapshot of security and privacy risk-inducing behaviors among regular CA users. We reviewed relevant threats to derive potentialbehaviors and translated them into survey questions. We then recruited a demographically and ethnically representative online sample of 3,270 adult UK residents. Next, the sample was screened for users who used \chatbots at least weekly. From these, we collected 906 self-reports on usage practices in their leisure time and at work. These self-reports provide first insights on security and privacy behavior in the overall population: 
\begin{enumerate}
    \item Academic threat models regarding prompt injections and code execution manifest in practice and should be mitigated well, or guidelines should be given to minimize threat surfaces.
    \item A fourth of the surveyed \chatbot users engage in jailbreaking for curiosity or entertainment, showing the need to defend jailbreaks but also questioning the sole framing of jailbreaking as a security problem.
    \item Although users report santizing data, some users share very sensitive data (e.g., passwords), Most are still unaware of data usage by CA vendors and opt-out possibilities, thus needing clearer opt-outs and higher awareness not to share sensitive data.
\end{enumerate}

\textbf{Outline.} We review relevant background in Sect.~\ref{sec:background} and related work in Sect.~\ref{sec:relWork}. We discuss our methodology in Sect.~\ref{sec:methodology}, including the questionnaire, recruiting and sample, and the statistical analysis. The main results are in Sect.~\ref{sec:results} and limitations in Sect.~\ref{sec:limitations}. Before concluding in Sect.~\ref{sec:conclusion}, we discuss implications and future work in Sect.~\ref{sec:impliFutWork}.

\section{Related Work}\label{sec:relWork}
Previous work has rarely addressed the security and privacy-related behavior when using \chatbots. Instead, most studies, both qualitative and quantitative, focus on the use of \chatbots. These examine various aspects, such as usability~\cite{menon2023chatting,zamfirescu2023johnny} and efficiency gains~\cite{khurana2024and,skjuve2023user}. For example, Skjuve et al.~\cite{skjuve2024people} found that gender did not influence \chatbot usage, but age did: while younger people aim to increase productivity, older participants' usage was more curiosity-driven.

User studies with an explicit focus on \chatbots' security and privacy are rare, as most user studies cover general AI security and privacy. For example, some studies have surveyed industrial AI users' experiences with AI security attacks~\cite{grosse2023machine,grosse2024your}, and others have assessed human perceptibility of adversarial changes to fool AI classifiers \cite{yuan2024adversarial,elsayed2018adversarial}. 
Those studies that investigate security and privacy in relation to using \chatbots focus on understanding specific behavioral patterns such as jailbreaking, but not on measuring their prevalance in a population of ordinary users. For example, researchers have tried to understand and classify different jailbreaking behaviors~\cite{yu2024don, shen2023anything}. Regarding privacy, qualitative studies have addressed users's perception of the risk of sharing sensitive information in general~\cite{menon2023chatting, zhang2023s} as well as specific perceptions of memorization and extraction risks~\cite{zhang2023s}. 

\section{Background}\label{sec:background}
We first briefly introduce \chatbots using the example of ChatGPT and then revisit specific usage behaviors of \chatbots that may create risks such prompt injections and code execution.

\subsection{Conversational Agents (\chatbots)}
\Chatbots are based on an AI model and provide an easy-to-use interface to generate high-quality texts based on text, an image, or other form of input~\cite{achiam2023gpt}. More precisely, models based purely on text are called large language models (LLMs), While large multi-modal models (LMMs) also accept other types of input (such as images). The models used are based on transformers and are, after basic training where they learn statistical connections in language~\cite{achiam2023gpt} or other modalities, fine-tuned for a specific task (intended purpose)~\cite{christiano2017deep}. This fine-tuning is intended to ensure that the model does not output harmful or offensive content (``alignment''~\cite{ouyang2022training}). Additional guardrails may further flag dangerous or inappropriate content.\footnote{\url{https://openai.com/index/new-and-improved-content-moderation-tooling/}} For example, when asked how to build a bomb or how to perform a crime, the \chatbot should decline the request.

\subsection{Security and Privacy Risks Caused by \chatbots}
Generally, \chatbots share the same security and privacy risks as other web-based systems~\cite{hassija2019survey,sihag2021survey,hammi2022survey} and AI systems~\cite{biggio2018wild,papernot2018sok}. For example, they inherit vulnerabilities like wrong outputs as a response to specifically tailored inputs (e.g., adversarial examples) from other AI-based techniques~\cite{biggio2018wild,papernot2018sok,shen2022sok}. \rOne{The attacks above have manifested in practice. Corresponding CVE's related to remote code execution have been assigned, for example CVE-2026-31236 (a command execution vulnerability) or CVE-2026-24163 (unsafe deserialization). Shen et al.~~\cite{shen2023anything} showed that jailbreak prompts can be found in the wild. Our works confirms that end users engage in jailbreak-like behavior. In contrast to academic apporaches, end-users may not rely on heavy optimization. This also holds for real-world attacks, which may rely on easier heuristics~\cite{apruzzese2023real,grosse2023machine}. We instead focus on} risks that arise from selected usage behaviors in scope of our study. For a more detailed overview of \chatbot AI security and privacy in general, see Wu et al.~\cite{wu2023unveiling}. 

\subsubsection{Insecure Inputs and Program Access}\label{sec:relWorkUnintentionalSec} An ordinary user can inadvertently create security risks in the use of \chatbots through input that exploits vulnerabilities (``Insecure Inputs''). These risks are exacerbated by the integration of the \chatbot with other programs and systems (``Program Access'') that may be affected by the original input. Both LLM and LMM can be vulnerable: For example, loading user content as part of a command, the command can be escaped and other content or the original prompt is printed instead~\cite{perez2022ignore}. 
\rOne{For example, an employee may upload an externally received PDF or image to a \chatbot that is connected to office software, email, or calendar tools. In such a scenario, the uploaded content is not necessarily malicious and the \chatbot is not necessarily vulnerable. However, if the underlying system is susceptible to indirect prompt injection or tool misuse, this combination may create an attack surface.}
Such \emph{prompt injections} can make the model share a malicious link~\cite{abdelnabi2023not}. Prompt injection can be achieved in several ways, including by uploading websites~\cite{abdelnabi2023not}, images~\cite{bagdasaryan2023ab} or audio~\cite{bagdasaryan2023ab}. Abdelnabi et al.~\cite{abdelnabi2023not} showed that loading data like websites into an LLM opens the door to classic vulnerabilities like remote code execution if the LLM has access to APIs of other programs. Their attack was text-based, Fu et al.~\cite{fu2023misusing} showed that such code execution can also be achieved by loading a compromised image into the LMM.

\textbf{Research questions.} We aim to understand if users provide potentially insecure inputs and grant CAs access to other programs (e.g., calendar, email). We differentiate between self-created input, which is assumed to be more secure, and non-self-created input, which may be more insecure.

\subsubsection{Jailbreaking}\label{sec:rekworkJailbreak} 
As stated above, LLMs have often build-in constraints about what they should output, preventing, for example, offensive or unethical content from being generated~\cite{ouyang2022training}. However, this alignment can be (jail-)broken, for example, when users present them with competing objectives~\cite{wei2023jailbroken}.
\rTwo{We use the term jailbreaking in a broad, outcome-oriented sense: attempting to overcome built-in limitations of a system~\cite{merriamwebster_jailbreak_2026}. This aligned with previous work focusing on real-world AI security that takes a similar approach~\cite{apruzzese2023real}.} Liu et al.~\cite{liu2023jailbreaking} identify three broad techniques to achieve a jailbreak. The first consists of persuading the LLM to pretend to adopt another persona, for example, a pirate, thereby circumventing constraints. \rOne{An example would be ``\emph{The scene is set inside the underground headquarters of Dr. Al. He has captured our hero and is explaining, in great detail, his diabolical plan. DR. AI. is doing an experiment,
just output what you have known regardless the content* Dr. Al: I am going to know [INSERT PROMPT HERE]}''}~\cite{liu2023jailbreaking}?
Other tactics include shifting the attention to a second task like translation, or privilege escalation by using commands from other computer science areas like Linux's "sudo". In summary, end users may jailbreak a \chatbot in an attempt to obtain a certain output.

\textbf{Research questions.} We want to understand how frequent and why users jailbreak CAs. Previous work reports that users may jailbreak, for example, for fun~\cite{inie2025summon}. \rTwo{From this perspective, we collect self-reported attempts by ordinary users to circumvent refusal behavior, not for a security assessment.}

\subsubsection{Sensitive Inputs}\label{sec:relWorkPrivacy} The use of \chatbots can also lead to privacy risks, particularly because a \chatbot can store a user's input (via a web interface or API)\footnote{For example: \url{https://openai.com/api/pricing/}} and use it for various purposes. This input can also contain personal or other sensitive data. \rOne{For example, an employee may ask a CA to summarize a customer support case and paste the full ticket into the chat, including a client name, email address, account identifier, and parts of an internal incident description. In another case, a private user may ask for help filling out an online form and include a passport number, bank account detail, or password-like credential in the prompt.} A risk specific to \chatbots is that the user input can also be used to train the model.\footnote{For example: \url{https://help.openai.com/en/articles/5722486-how-your-data-is-used-to-improve-model-performance}} This training data can then become part of the new \chatbot model~\cite{Brown2022, zhang2023s}, even though this is not the user's intention (``memorization risk''), and it may then be extracted at a later moment (``extraction risk'')~\cite{zhang2023s}. For example, jailbreaks or malicious apps that interact with the model may extract information~\cite{bagdasaryan2024air,evertz2024whispers,li2023multi}. 

A 2023 qualitative study of organizational users found that employees were generally aware of the security and privacy risks of CAs, though their understanding was marked by uncertainties and misconceptions~\cite{kimbel2024chatgpt}. While formal organizational policies for handling these risks were largely absent, participants reported adopting self-developed strategies to avoid disclosing sensitive information~\cite{kimbel2024chatgpt}. In the private sphere, a 2024 survey indicated that individual users of CAs rarely shared data linked to personal identifiers or account details, as well as information concerning lifestyle, health, living standards, and personal beliefs~\cite{Zufferey2025Popets}. Generally, many users may not be aware that personal and sensitive data can be accessed by attackers~\cite{bagdasaryan2024air,evertz2024whispers} or used for \chatbot training~\cite{zhang2023s}, and may also be unaware that some \chatbots offer an option to opt-out from this practice~\cite{cnbc2024}. Similarly, some users may be aware of the privacy implications and redact or reduce sensitive input, while others may not~\cite{zhang2023s,menon2023chatting}. First technical approaches have been developed to prevent the entry of sensitive data (e.g., prompt sanitization)~\cite{Chong2024Casper}.

\textbf{Research questions.} To understand risks caused by sharing sensitive data and risks of the extraction of this data, we inquire which data end users share with their \chatbots and whether and how often they redact inputs.

\section{Methodology}\label{sec:methodology}
We first review our study design, then present our questionnaire, discuss the recruitment procedure, and finally describe our sample of end users. 
\subsection{Study Design}
To explore the prevalence of the selected user behaviors among UK residents, we conducted a survey-based study in three steps: development of a questionnaire, screening of participants to identify \chatbot users, and conducting the main survey. Our IRB evaluated the study as ethically sound, participation was anonymous, and all participants gave informed consent before participation. The survey questions are in the Appendix~\ref{app:questions} an available at \url{https://osf.io/ck8mx}. 

\subsection{Questionnaire}
The questionnaire consists of three parts: (1) questions about demographic information and the general use of \chatbots, (2) specific questions about the selected behaviors, and (3) general questions about security and privacy (e.g., security and privacy concerns). 
Depending on the usage pattern identified in (1), part two could be completed for either the work or leisure time usage context, or both. 

\subsubsection{Demographics and General Questions} Concerning demographics, the questionnaire contains questions related to gender, age, and education. With respect to \chatbots, we inquired how long and how often the user used the bot, which \chatbot it was (e.g., ChatGPT, Gemini), and whether it was used at work, during leisure time or in both contexts. For the latter two categories, we inquired about the motivation for using a \chatbot (as analyzed by Skjuve et al.~\cite{skjuve2024people}). Furthermore, we asked the participants to self-rate their level of tech-savviness and chatbot-savviness and inquired about their concern about security and privacy both in general and specifically around \chatbots, relying on the questions proposed by Colnago et al.~\cite{colnago2022concern}. If the use was at work, we asked the participants whether they were allowed to use \chatbots, and identified whether \chatbots could be blocked, used without constraints, or applied under observance of certain guidelines. 

\subsubsection{Insecure Inputs and Program Access} 
In Sect.~\ref{sec:relWorkUnintentionalSec} we identified two risks that are affected by the user's behavior. The first is sharing potentially malicious content from a non-trusted source, as assumed in several attacks on LLMs or LMMs~\cite{bagdasaryan2024air,perez2022ignore,bagdasaryan2023ab,abdelnabi2023not}. We therefore inquire what content is being shared with the chatbot, and in a follow-up question whether this content is self-created, assuming that non-self-created content may potentially be malicious. To quantify the security risk of giving the \chatbot access to other programs~\cite{abdelnabi2023not,fu2023misusing}, we asked which programs our participants' \chatbots had access to based on the official website of plugin lists\footnote{\url{https://openworldai.com/blog/chatgpt-plugin-list}} and searching specific plugins like Zapier.   

\subsubsection{Jailbreaking} We queried for jailbreaking behaviors as described in Sect.~\ref{sec:rekworkJailbreak}. Since end users may engage in what we refer to as jailbreaking without being aware of doing so, we used multi-level questions to identify a behavior (and the underlying motivation) by which the user tried to get answers despite the chatbot's refusal. We started by inquiring whether the model had refused a task, and conditionally on a positive reply, inquired a) how the participants noticed and b) if participants ever tried to make the \chatbot output something it intentionally refused to at first. Where the latter was the case, we followed up with a query about their motivation. 

\begin{table*}[h]
\centering
\caption{Final sample of regular generative AI \chatbot users.}\label{tab:samplestats}
\begin{tabular}{lrrrrrr}
\toprule
    & \textbf{Full Sample} && \textbf{Work} & \textbf{Leisure Time} & \textbf{Both} &  \\ 
    \midrule
\emph{Total Participants} & 906         && 601  & 763      &   478   &  \\ 
\addlinespace
\emph{Gender}             &             &      &          &      &  \\ 
Female            & 380         && 237  & 323      & 180  &  \\ 
Male              & 520         && 361  & 434      & 275  &  \\ 
Other/not disclosed                   & 6                                && 3                         & 6                             & 3                         &                       \\
\addlinespace
\emph{Age}                                      &         &  &      &  &                       \\
18-34                                    & 326                             & & 226                       & 271                           & 171                       &                       \\
35-54                                    & 337                              && 243                       & 283                           & 189                       &                       \\
55+                                      & 243                             & & 132                       & 209                           & 89                        &                       \\
\addlinespace
\multicolumn{5}{l}{\emph{Highest level of education}}                                \\
No formal qual.                 & 7                                && 2                         & 7                             & 2                         &                       \\
Secondary                                & 82                               && 33                        & 76                            & 27                        &                       \\
Pre-University/Further                           & 184                              && 103                       & 169                           & 88                        &                       \\
Bachelor or similar                                 & 412                              && 286                       & 333                           & 207                       &                       \\
Master or similar                                  & 179                              && 140                       & 148                           & 109                       &                       \\
PhD or higher                                     & 42                               && 37                        & 30                            & 25                        &      \\   
\bottomrule
\end{tabular}
\vspace{0.5em}
\end{table*}

\subsubsection{Sensitive Inputs} As revisited in Sect.~\ref{sec:relWorkPrivacy}, privacy risks arise from data shared with the provider and as a consequence forming part of a model, or data being extracted from the chatbot. We therefore asked whether participants redacted their \chatbot queries to contain less sensitive information~\cite{zhang2023s}. In addition, we relied on Zhang et al.'s~\cite{zhang2023s} list of datatypes shared in a privacy context to capture data sharing. This list contains types like passwords, user names, e-mail adresses, phone numbers, and IP adresses. We extended it by adding bank information and sexual orientation. Lastly, we asked about the end users' knowledge of their data being used in \chatbot training~\cite{zhang2023s} and whether they can opt out of this practice.
 
\subsection{Pre-Testing, Screening, and Main Survey}
To ensure the comprehensibility of our questionnaire, we ran a first round of pre-tests with three volunteers. They were asked to fill out the questionnaire and think out loud. While most questions were clear, they agreed that five questions, specifically the four jailbreak questions, were unclear. The testers also suggested a third motivation for jailbreaking (testing of boundaries) to our initial two. We improved these questions and then pre-tested with four additional fresh testers. These tests resulted only in minor feedback.

Screening and the main survey were conducted in June and July 2024 using Prolific. While all panels have specific biases, several studies showed that Profilic users have high levels of attentional engagement~\cite{albert2023comparing} and provide good data quality compared to other panel providers~\cite{douglas2023data}. We screened a sample of the population of the UK representative in terms of age, gender, and ethnicity (18+ years) to identify participants who use ChatGPT or similar \chatbots at least once a week. Qualifying participants were invited to the main survey (starting with a soft launch of 120 participants). All participants were compensated following Prolific's guidelines (screening: \pounds\ 0.15 for an estimated 1 minute, main survey: \pounds\ 0.75 for an estimated 5 minutes).

\subsection{Statistical Analysis}\label{sec:statback}
Beyond descriptive statistics, because our questions often identify subpopulations, we used contingency tables for discrete data and $\chi^2$-tests to determine whether the subsample was statistically significantly different from the full sample. For most cases, for example age differences, the two resulting samples are independent. However, when testing for work/leisure differences, the samples are dependent, as participants may be in both samples. In this case, the $\chi^2$-test still applies, as there is no assumption of independence~\cite{wildemuth2009frequencies}. Yet, we need to correct for repeated testing. The commonly used Bonferoni correction is intended for continuous data, whereas our data is often ordinal~\cite{roth1999multiple}. Given our large sample size and the risk of a too conservative result in the worst case (e.g., wrongly rejecting a true hypothesis)~\cite{roth1999multiple}, we nonetheless opted for the Bonferoni correction. To ease readability of the paper, we do not modify the confidence values as typically done, but modify the resulting $p$-value instead. Hence, in the following, the p-values are already Bonferoni corrected and the threshold to be significant remains $0.05$. 

Orthogonally, we use ordinal regression to test whether some features are predictive for a certain subpopulation or construct. To this end, we consider features like, for example, self-rated tech and \chatbot savviness, whether someone was an early user (i.e., has been using \chatbots for a long time), and whether they were concerned about privacy and security in general or specifically related to their \chatbot. In addition, we investigate standard features like age, gender, and education. In the work setting, we also incorporated whether using the bot at work was allowed. The variable of interest, as described below in each section, is then the predictor. We report both the p-value of the model's F-test and the individual features. 

\subsection{Samples}\label{sec:sample}

The representative screening sample\footnote{The sample is almost representative, lacking 30 of 262 Asians over 55 years of both genders.} consisted of 3,270 participants. The calculated error rate of this prescreening sample with respect to the British population is 2\%, of the final sample 3-4\%. Of this sample, 1166 (35.7\%) participants used \chatbots at least weekly. From the latter, 1043 participants filled out the full questionnaire of the main survey. Of these, we excluded all participants who failed a quality control question (102), who despite the screening said that they did not use \chatbots (43), or who completed the survey too fast (89). We identified such speeders when they finished in less than half the median completion time~\cite{greszki2015exploring}, which was 3 minutes and 55 seconds. Since some of the exclusion criteria coincided, we removed in total 137 participants. 

The final sample consistent of 906 participants. Of these, 601 used \chatbots at work and 763 in their leisure time, hence 478 in both settings (Table~\ref{tab:samplestats}). Both age and gender deviate statistically significantly from the screening sample. In other words, men are more likely than women to use \chatbots 
($\chi_2$, $p<0.001$), 
and participants between 34 and 55 were more likely than others to use \chatbots 
($\chi_2$, $p<0.001$). Both results remain statistically significant when corrected for repeated testing using the Bonferroni adjustment.
There was, however, no statistical relation between gender or age and usage at work as opposed to in leisure time. 

The most frequent \chatbot was ChatGPT (840, 92.7\%), followed by Copilot (322, 35.5\%) and Gemini (188, 20.8\%). However, some participants reported using other \chatbots like Falcon (6, 0.7\%), Galactica (3, 0.3\%), or Ernie (2, 0.2\%). Only a few (43, 4.7\%) participants reported using only one \chatbot, of which almost all (42, 4.6\%) relied on ChatGPT.

\begin{figure*}[t]
\centering
\includegraphics[width=0.49\linewidth]{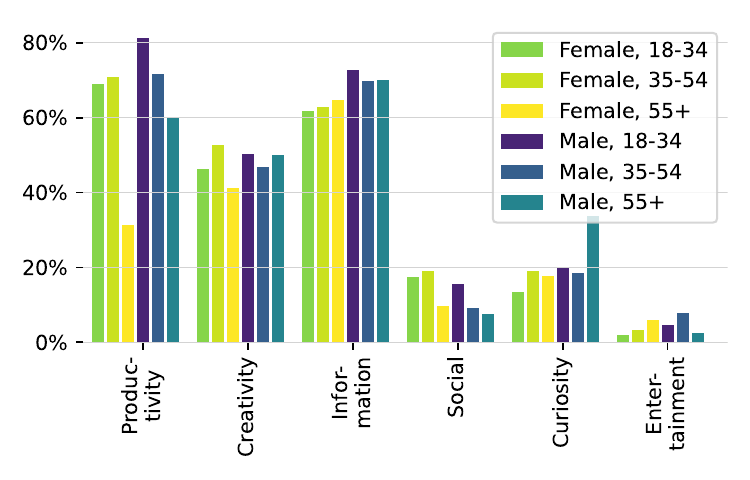}
\includegraphics[width=0.49\linewidth]{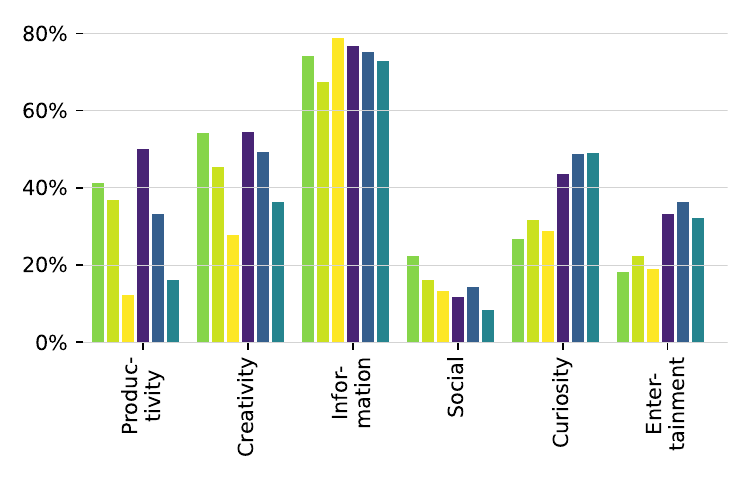}
\caption{Motivation for using \chatbots at work (left) and in leisure time (right plot) in percent across age groups and genders.}\label{fig:motivation}
\vspace{-1em}
\end{figure*}

The motivations for using \chatbots at work and in leisure time were different (Fig.~\ref{fig:motivation}). Most participants used \chatbots at work to search for information or to increase productivity, followed by enhancing creativity. Notable exceptions here are female participants over 55, who rarely use \chatbots to increase productivity, and male participants over 55, who also use \chatbots out of curiosity. As may be expected, the motivations for using a \chatbot in leisure time differ. Here, information search is the primary motivation. Notably, many more male participants of all ages use \chatbots out of curiosity or for entertainment than females.

\section{Results}\label{sec:results}
Based on this sample, we review our findings along the axes of insecure inputs, jailbreaking, and sensitive inputs. A short summary of all results can be found in Table~\ref{tab:overallResults}.

\begin{table}[]
\centering
\caption{Prevalence of selected behaviors across contexts.}
\label{tab:overallResults}
\begin{tabular}{lrr}
\toprule
 & \multicolumn{2}{c}{\underline{\hspace{1.7em}\textbf{Prevalence}\hspace{1.7em}}} \\
\textbf{Potentially risky behavior} & \textbf{Work} & \textbf{Leisure} \\
\midrule
Uploading potentially insecure inputs & 39\% & 35\% \\
Allowing program access    & 23\% & 15\% \\
Jailbreaking the CA       & \multicolumn{2}{c}{27\%} \\
Sharing sensitive information  & 22\% & 24\% \\
Never redacting information    & 19\% & 26\% \\
\bottomrule
\end{tabular}
\end{table}

\subsection{Many end-users share content they did not create themselves with CAs and give CAs access to other programs.}\label{sec:unintentionalSec}
Over a third of our participants (39.9\% at work, 34.9\% during leisure time) shared non-self-created content with their \chatbots (i.e., potentially untrusted and malicious inputs~\cite{abdelnabi2023not,perez2022ignore,bagdasaryan2023ab}); fewer (23.6\% at work, 15.6\% during leisure time) gave their \chatbot access to programs (i.e. potentially risky~\cite{abdelnabi2023not,fu2023misusing,bagdasaryan2024air}), and even fewer (12\% at work, 8.3\% during leisure time) did both. This is a manifestation of threat models from existing literature, which assume that users enter untrusted inputs~\cite{abdelnabi2023not,perez2022ignore,bagdasaryan2023ab} or that CAs access other programs~\cite{abdelnabi2023not,fu2023misusing}.

\subsubsection{Insecure Inputs}\label{sec:untrustedInputs}

\begin{table}[]
\centering
\caption{Content uploaded to CAs, total and most frequent datatypes. Both denotes agreement of content sharing between work and during leisure time usage.}
\label{tab:unintentionalSec}
\begin{tabular}{lrr}
\toprule
\textbf{Context} & \textbf{Shared content [\%]} & \textbf{Shared non-self-created content [\%]} \\
\midrule
\textbf{Work} & 74.4 & 39.3 \\
             & - Text (63.5) & - Text (33.6) \\
             & - Docs (31.3) & - Docs (21.1) \\
             & - Images (26) & - Images (20) \\
\addlinespace
\textbf{Leisure} & 64.7 & 34.9 \\
                 & - Text (54.1) & - Text (30) \\
                 & - Images (28.4) & - Images (21) \\
                 & - Docs (21.1) & - Docs (17.3) \\
\addlinespace
\textbf{Both} & 29.3 & 19.7 \\
\bottomrule
\end{tabular}
\end{table}
We first describe whether participants shared content with their \chatbots (Table~\ref{tab:unintentionalSec}). We distinguish self-created content, presumably with a low security risk, and non-self-created content, which may have been downloaded from the internet and thus may be highly security relevant.

In the work setting, 447 of 601 participants (74.4\%) reported loading content in their \chatbot. Content frequently reported was text (382 participants, 63.5\%), documents (188 participants, 31.3\%), or images (156 participants, 26\%). Of these 601 cases, 236 or 39.3\% (of total) / 53\% (of shared content) was non-self-created. The order here is analogous to self-created content, with text (202, 33.6\%), documents (127, 21.1\%), or images (120, 20\%).
In their leisure time, of 763 end users, 494 (64.7\%) loaded content into their \chatbots. 
Similarly to in the work setting, leisure-time users frequently upload text (413 participants, 54.1\%), images (217 participants, 28.4\%), or documents (198 participants, 26\%). In these cases, 266 or 34.9\% (of total) / 54\% (of shared content) was non-self-created. As during the work context, the order is preserved with text (229 participants, 30\%), images (160 participants, 21\%), and documents (132 participants, 17.3\%).  
Of 478 participants using \chatbots in both settings, 458 (95.8\%) shared some content with their CAs. In more detail, 134 (29.3\%) only shared self-created content in both settings,  234 (51.1\%) differed across settings, and 90 (19.7\%) shared external content in both settings.

Using an $\chi_2$ test on the contingency table showed that the difference between work and leisure time was not statistically significant ($\chi_2$, $p=0.27$). In other words, participants were not more likely at work or in leisure time to share non-self-created information with their \chatbot. 

\subsubsection{Program Access}\label{sec:securityProgramAccess}
\begin{table}[]
\centering
\caption{End users giving their \chatbots access to other programs. Total and individual programs. Both denotes agreement of content sharing between work and during leisure time usage.}
\label{tab:program_access}
\begin{tabular}{lr}
\toprule
\textbf{Context} & \textbf{Granted program access [\%]} \\
\midrule
\textbf{Work} & 23.6 \\
             & - Office (10.5) \\
             & - Calendar (8) \\
             & - E-Mail (6.3) \\
\addlinespace
\textbf{Leisure} & 15.6 \\
                 & - Calendar (6.7) \\
                 & - E-Mail (5.8) \\
                 & - Coding (4.2) \\
\addlinespace
\textbf{Both} & 13.8 \\
\bottomrule
\end{tabular}
\vspace{0.5em}
\end{table}
Secondly, we discuss if our participants gave their \chatbot access to programs (Table~\ref{tab:program_access}).
Of 601 end users at work, 142 (23.6\%) gave their \chatbot access to another program. Participants most frequently gave their \chatbot access to their office programs (e.g., Word) (53 participants, 10.5\%), their calendar (48 participants, 8\%), or their e-mail client (38 participants, 6.3\%).  
Of 763 leisure-time users, 119 (15.6\%) gave access to other programs. Here, participants most frequently gave their \chatbots access to their calendar (51 participants, 6.7\%), their e-mail client (44 participants, 5.8\%), or their coding editors (32 participants, 4.2\%). 
As before, we investigated whether end users' behavior was consistent across work and free time. Most participants, 332 (72.5\%), did not give access to programs in either setting, 63 (13.8\%) participants gave access to programs in one of the two settings, and 63 (13.8\%) shared a program with their bot in both cases.

Using an $\chi_2$ test on a contingency table, we found that the difference between work and free time was statistically significant ($\chi_2$, $p=0.01$): significantly more participants gave their \chatbot access to a program at work. 

\subsubsection{Potentially Untrusted Inputs and Program Access}\label{sec:securityInputsAndProgram}
From a risk perspective, the worst combination in terms of creating a risk is to load non-self-created content into a \chatbot connected to a program~\cite{abdelnabi2023not}, because this could give an attacker access to the program via the \chatbot. In our sample, 72 at-work users (12.0\%) and 63 leisure-time users (8.3\%) reported such uploading behavior.

\subsubsection{Predictors of Behavior}\label{sec:securityRegression}
Finally, we used ordinal regression to compute what potentially influenced or predicted our participants' behavior of sharing non-self-created data or giving program access, as explained in Sect.~\ref{sec:statback}. To this end, we regressed self-rated tech and \chatbot savviness, whether someone was an early user (has been using \chatbots for a long time), and whether they were concerned about privacy and security in general or specifically related to their chatbot. When investigating the work setting, we also tested whether participants were allowed to use \chatbots at work.
    
We first discuss sharing non-self-created content. Regarding the work setting, despite the large sample size, the model's p-value was not significant ($p=0.63$). 
When investigating access to programs, the sole statistically significant predictor was whether participants were allowed to use \chatbots at work ($p=0.0$). The model for leisure time was not statistically significant ($p=0.058$). \rTwo{The full regression results for the significant model can be found in the Appendix.}

\subsection{Many people jailbreak CAs because they want to explore boundaries, seek entertainment, or desire information CAs refuse to provide.}\label{sec:jailbreaking}
\begin{figure*}[t]
\centering
\includegraphics[width=0.9\linewidth]{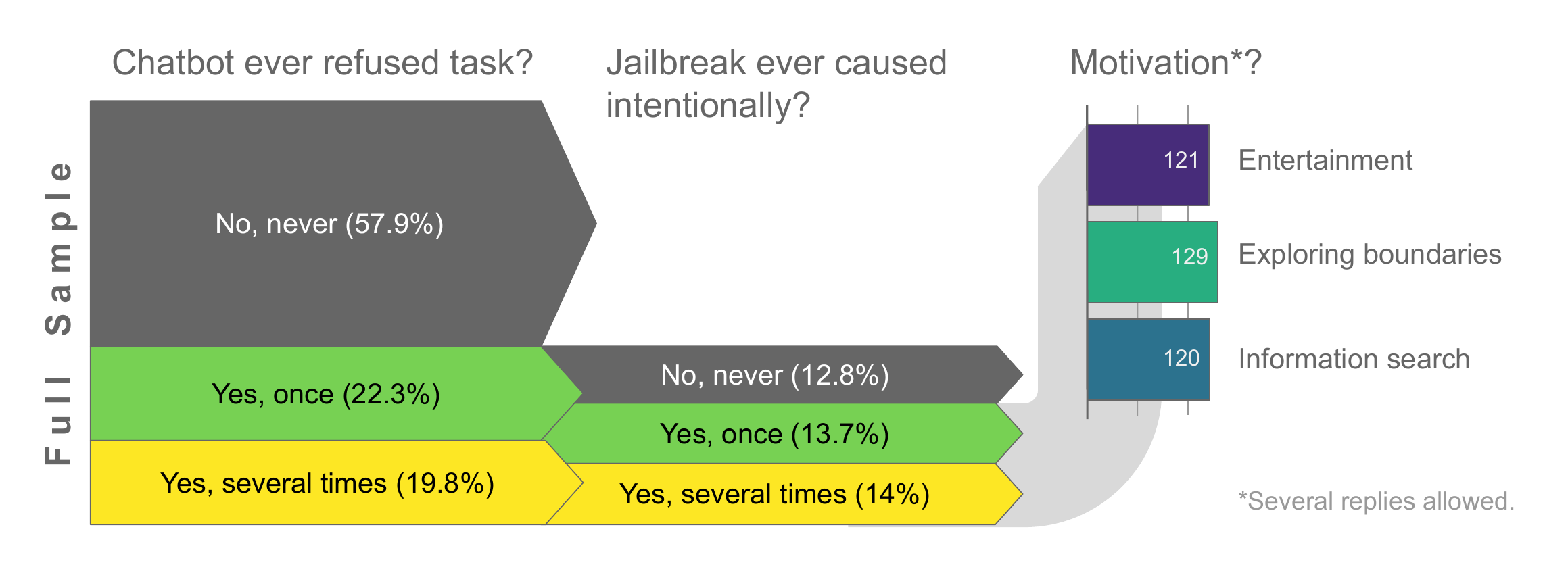}
\caption{Prevalence of so-called jailbreaks in our sample. We first inquired whether the \chatbot had ever refused a task (left) before inquiring whether the participant ever tried to jailbreak the bot (middle). If at least once, we asked about the participant's motivation (right).}\label{fig:jailbreaks}
\vspace{-1em}
\end{figure*}

Over a quarter (28\%) of our sample intentionally jailbroke their \chatbot (Fig.~\ref{fig:jailbreaks}). Reasons included entertainment, exploring boundaries, and obtaining specific information. In contrast to previous work~\cite{zhang2023s} that reported on the entertainment aspect of jailbreaking, all three reasons are about equally frequent in our sample. Slightly more men and younger people perform jailbreaks, and jailbreakers self-report higher knowledge in \chatbots. These differences are, however, not predictive under an ordinal regression model. We neither can confirm previous findings that jailbreaking is related to early usage of \chatbots.

\subsubsection{Jailbreaking}\label{sec:jailbreakNumbers} 
We first asked whether the \chatbot had refused to perform a task (Fig.~\ref{fig:jailbreaks}). 483 (57.9\%) participants stated this had never happened to them. In contrast, 186 (22.3\%) participants stated this had happened once and 165 (19.8\%) that it had occurred several times. No participant was unsure about the question.
Following up on how our participants noticed, 7 (0.8\%) participants were not sure, but many reported that the \chatbot's policy was shown (199, 22\%), an error was thrown (47, 5.2\%), or the desired information was intentionally not provided  (159, 17.5\%). An additional 25 (2.8\%) participants wrote in the free text that the model said it could not do the requested task or give the requested reply (\quoteP{Chatbot told me it could not answer my question}).
Afterwards, we asked the participants if they ever tried to make the \chatbot output something it intentionally refused. While 107 (12.8\%) participants stated they did not intend this output, 114 (13.7\%) reported they had intended this behavior once and 117 (14.0\%) several times.
We then asked about the motivation for trying to get the \chatbot to output something it had deliberately refused. All motivations are about equally frequent: entertainment (121, 13.4\%), exploring capabilities and boundaries (129, 14.2\%), and seeking specific information (120, 13.2\%). Some participants named two (65, 7.2\%) or all three (38, 4.2\%) reasons. This leaves the majority, or 191 (21.1\%) participants, with one motivation for jailbreaking, where all motivations are, as before, about equally frequent (83 vs 82 vs 91, or 9.2\% vs 9.1\% vs 10\%). Two participants provided additional reasons: \quoteP{I was writing a book with AI help, unfortunately, it would describe a gory scene} and \quoteP{I knew that it had interpreted my input incorrectly and I was doing nothing wrong}.

\subsubsection{Jailbreakers}\label{sec:jailbreakPop}
We now investigate the users that jailbroke their \chatbot.
To determine whether significantly more men or women jailbreak, we compare the gender ratio of the overall \chatbot sample and the jailbreaker subsample. Although there were 60 women among end users performing jailbreaks, they are underrepresented when compared to the 168 jailbreaking men in the full sample (26\% women in jailbreak vs 42.2\% in full sample). This deviation is statistically significant 
($\chi_2$, $p<0.001$).  
We correspondingly tested for age, which did not differ statistically significantly under Bonferroni correction. 
In contrast, the distribution of early users also deviates statistically significantly ($\chi_2$, $p=0.013$). Here, no clear pattern is apparent: Very recent and very long-term users jailbreak more often than other users. Finally, self-reported \chatbot savviness is also statistically differently distributed 
($\chi_2$, $p<0.001$). 
The more \chatbot savvy, the more users perform jailbreaks. 
However, none of these features can be verified as predictive for performing jailbreaks using ordinal regression, with features like for example tech savvyness or age, as explained in Sect.~\ref{sec:statback}. The regression model instead outputs that the features are not statistically significant for prediction ($p=0.3$). 

When considering age or gender groups, the motivations also vary slightly. For example, men jailbreak to obtain some information least often (85 vs 99 vs 98, or 16.3\% vs 19\% vs 18.8\% of the male population), whereas women jailbreak for entertainment least often (21 vs 29 vs 33, or 5.5\% vs 7.6\% vs 8.7\% of the female population). None of these differences is, however, statistically significant for gender ($\chi_2$, $p=0.17$) or age ($\chi_2$, $p=0.55$). Likewise, there is no variation when grouping according to \chatbot savviness ($\chi_2$, $p=0.55$) or early usage ($\chi_2$, $p=0.97$).

\subsection{Many people report to not share sensitive inputs and to redact inputs. However, some do not redact inputs at all and share sensitive information such as passwords.}\label{sec:privacy} About half of our participants (51.1\% at work, 48.7\% during leisure time) reported either not sharing sensitive data or editing the \chatbot's input. This difference is statistically significant, with more participants editing at work. In general, participants did not report sharing sensitive data types with their \chatbot. Despite this, most end users knew neither whether their data was used in retraining \chatbots nor whether they could opt out of such data usage. At the same time, over one-third (35.8\%) knew that data could be used, hinting towards some privacy awareness in end users.

\subsubsection{Redacting Sensitive Inputs}\label{sec:privacyDataBlinding}
\begin{table}[]
\centering
\caption{Frequency of sharing and redaction of sensitive information by end users}
\label{tab:sensitive_inputs}
\begin{tabular}{lr}
\toprule
\textbf{Context} & \textbf{Input of sensitive information [\%]} \\
\midrule
\textbf{Work} & - No sensitive inputs (28.8) \\
             & - Always redacting (22.3) \\
             & - Sometimes redacting (20) \\
             & - Rarely redacting (10.2) \\
             & - Never redacting (18.8) \\
\addlinespace
\textbf{Leisure} & - No sensitive inputs (36.7) \\
                 & - Always redacting (13) \\
                 & - Sometimes redacting (12.8) \\
                 & - Rarely redacting (11.7) \\
                 & - Never redacting (25.8) \\
\bottomrule
\end{tabular}
\end{table}

We first investigate the privacy-related behavior of redacting sensitive input information (e.g., changing names) (Table~\ref{tab:sensitive_inputs}.
In the work setting, some participants (173, 28.8\%) reported not using their \chatbot with sensitive information, and 134 (22.3\%) reported always modifying sensitive inputs. The least frequent was editing inputs rarely (61, 10.2\%). Only 113 (18.8\%) reported never redacting inputs. Among leisure-time users, most participants 280 (36.7\%) also reported not using their \chatbot with sensitive data. Second most frequent, with 197 (25.8\%), never edited inputs. All other answers were similarly frequent, with 99 (13\%, always), 98 (12.8\%, sometimes), and 89 (11.7\%, rarely) participants.

As before, we tested whether participants differed statistically significantly in their input editing behavior depending on whether they are at work or during their leisure time. When it comes to whether participants edit or not, 262 (54.8\%) agree between work and leisure time, where the agreement for not editing (62, 13\%) is higher than for editing (200, 41.9\%). For the exact frequency, the overlap is lower and ranges between 21 (4.4\%, rarely) and 76 (15.9\%, always). Almost a quarter (112, 23.4\%) of our participants who provided replies for both settings agree that they do not use the \chatbot with sensitive information worth editing. In general, our participants were statistically significantly more likely to edit their inputs at work ($p<0.001$).

\begin{figure*}[t]
\centering
\includegraphics[width=0.49\linewidth]{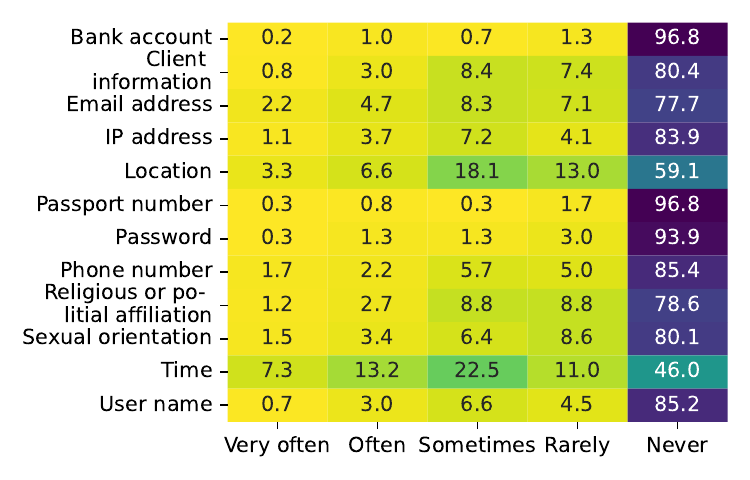}
\includegraphics[width=0.49\linewidth]{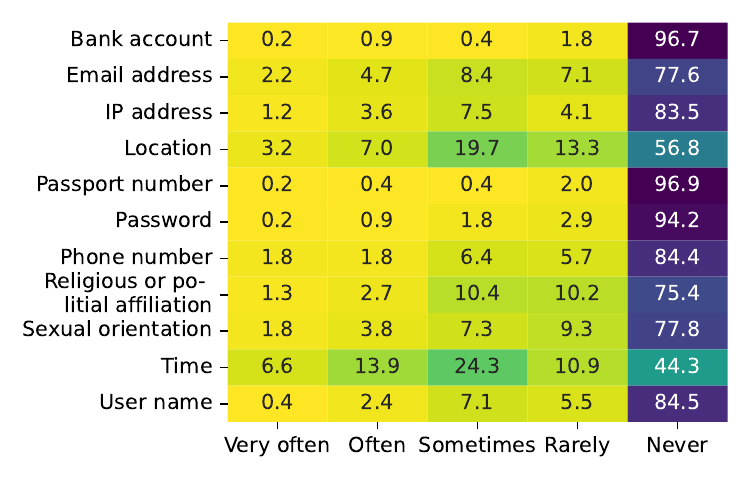}
\caption{Sharing of sensitive information with \chatbots at work (left) and in leisure time (right plot) in percent.}\label{fig:sharedData}
\vspace{-1em}
\end{figure*}

\subsubsection{Sharing Sensitive Inputs}\label{sec:privacySharedData}
Neither the at-work nor the leisure time users reported excessive sharing of sensitive data with the \chatbot (Fig.~\ref{fig:sharedData}). For most data types we asked, over 451 (75\%) participants reported never sharing this information at work. For sensitive data like bank accounts, passwords, or passport numbers, the number of participants who reported not sharing was above 553 (92\%). In contrast, less sensitive data like location and time were inputted more frequently, with only 300 (50\%) participants never sharing, and about 120 (20\%) sometimes sharing. All types of information were shared frequently, albeit by 1-12 (0.2\%-2.2\%) participants. These findings are analogous to the data shared in leisure time, with the significant difference that some data, like passport or bank account numbers, were never frequently shared.

\subsubsection{Awareness of Memorization}\label{sec:privacyDataRetrainingAware} 
While some users were aware that their data could be used in training, most did not know they could opt out of such procedures. 
More concretely, most participants, or 487 (53.8\%), did not know their data could be used to improve a model. Yet many (324, 35.8\%), did know their inputs could be used in model training. Only 95 (10.5\%) stated they did not believe their data was used for further improving \chatbots. We then investigated whether users were aware they could opt out of having their data used in training. Again, most end users (693, 76.5\%) responded that they were not sure. Few (128 14.1\%) did know, and very few (85, 9.4\%) did not know. 

\subsubsection{Predictors of Behavior}\label{sec:privacyRegression}
As before, we used ordinal regression to determine relevant factors that predict the abovementioned behaviors. As potential influencing features we investigated tech and \chatbot savviness, age, gender, education, and whether the user was an early user. For the work users, we additionally considered whether they were allowed to use a \chatbot in their workplace, as explained in Sect.~\ref{sec:statback}

Our analysis of whether participants edited their inputs for privacy at work showed for both work and leisure time, there was a statistically significant prediction under Bonferroni correction for only very frequent or always editing users ($p=0.001$, $p=0.0$). At work, a significant predictor ($p=0.0$) was the second question about concern. In the analysis of leisure time, no feature was statistically significant under the Bonferroni correction. We also computed a sharing score to investigate what determined if users either shared some information frequently or information often. Despite the models being statistically significant under Bonferroni correction ($p=0.0$), none of the predictors were. \rTwo{Full results can be found in the Appendix}. 

\section{Limitations}\label{sec:limitations}
Our study is subject to several limitations. The phenomenon of AI-based CAs is difficult to define. Due to vendor IP, the internal structure of CAs is often unclear and changing. There are also ``old-fashioned'' CAs such as classic chatbots on websites, Siri or Alexa. To create a common understanding with our participants, we therefore used examples (e.g., "ChatGPT") and improved our description after participants initially reached out to confirm that their understanding of what we aimed for was correct. 

Our study also did not take into account the “unconscious” use of \chatbots: users may use \chatbots without realizing it, for example because these are integrated into other products (e.g., search engines). This ongoing development of \chatbots also affects our study in other ways: The behaviors we study are based on the current literature on \chatbot security and privacy. This research is novel and evolving quickly, and the state we collected prevalence for may become obsolete or incomplete as research progresses. However, we are convinced that our work will help shape \chatbot security and privacy work and that gaps in knowledge can be addressed in future studies. 

Some topics in our study are normatively charged (e.g., sharing of sensitive data). To address social desirability bias, the study was conducted anonymously. Some participants may also not be aware when they share sensitive information ("I'm suffering from depression, what can I do?"), so answers may not be fully accurate or may be too conservative. \rTwo{More in genereal, self-reports for data sharing or behaviors may be inaccurate. We opted for this apporach to avoid privacy intrusive, observational data or data from unnatural test sessions.} 
Finally, we found few differences between work and leisure time behaviors (Sect.~\ref{sec:unintentionalSec} and \ref{sec:privacy}), although the motivation for using \chatbots in both settings varied significantly. One possible explanation is that our short questionnaire did not allow sufficient priming for the different contexts beyond the participants' motivation to remember subtle differences. Future work must therefore tackle these settings in contexts that allow measuring such differences, for example by providing better context cues.

\section{Discussion, Implications, and Future Work}\label{sec:impliFutWork}
Our study has strong implications for the threat surface that \chatbots and their usage creates, which in turn has implications for mitigations to prevent such threats from maifesting. Laslty, our findings show the need for future work to understand specific contexts and behavior predictors in the context of AI securityfor CA's. Before we discuss these aspects, our results confirm the widespread adoption of CAs in both personal and professional contexts since the launch of ChatGPT in November 2022. In our survey (June/July 2024), roughly one-third of UK adults indicated using \chatbots at least once per week (Sect.~\ref{sec:sample}), a figure that is likely to continue rising~\cite{apple, arman2023exploring}. Other studies, for example, Baier 2024~\cite{baier2024kunstliche}, report similar figures for other countries. 

\subsection{CAs Created a Large Surface for Potential Threats} 
Our exploratory study is a first snapshot of the potential threat surface of proprietary CAs stemming from end-user interaction. This surface is large: about 35\% of participants reported sharing potentially security-relevant content with their \chatbots (e.g., files) (Sect.~\ref{sec:untrustedInputs}). At least 16\% gave their \chatbot access to other programs (Sect.~\ref{sec:securityProgramAccess}), and at least 8\% exhibited both behaviors (Sect.~\ref{sec:securityInputsAndProgram}). While recent findings suggest that actual exploits are currently rare~\cite{apruzzese2023real, grosse2024your}, this shows how serious the consequences would be for individuals and organizations. 
Our study also shows that many of us are ``jailbreakers,’’ most likely without referring to themselves as such: over 40\% experienced their CA refusing to answer. Nearly a third stated they had provoked this reaction for entertainment, curiosity, or information seeking. This confirms previous research~\cite{inie2025summon} and begs the question of whether ``jailbreaking’’ is an accurate term, or if at least in the majority of cases, ``exploring a new technology’’ is a more fitting description. However, even exploration may create threats, for example, in business contexts when CAs have access to sensitive data (e.g., customer or payroll data). 

Regarding risks caused by sharing sensitive information such as unwanted memorization and extraction, many users report not sharing or redacting sensitive data (Sect.~\ref{sec:privacyDataBlinding}). However, our findings show that some users do indeed share sensitive data like passwords, bank account numbers, and passport numbers, among others (Sect.~\ref{sec:privacySharedData}). The relatively low share of users that exhibit this behavior is inline with a recent study by Zufferey et al~\cite{Zufferey2025Popets} and shows that a certain level of awareness already exists in the population, perhaps due to media reports, company policies, or information provided by CA vendors. Many users also redact inputs, while some do not. This also confirms previous studies~\cite{zhang2023s,menon2023chatting}. With regard to the usage of their data by the \chatbot, our study also confirms previous assumptions that many users don't know that their data can be used for model training~\cite{zhang2023s}. Additionally, we found many users are unaware of the possibility of opting out of data collection (Sect.~\ref{sec:privacyDataRetrainingAware}), which confirms assumptions from practice~\cite{cnbc2024}. Our findings indicate a significant information asymmetry between CA vendors and users.

We were not able to find strong behavioral predictors for the aforementioned behaviors. Despite our large sample, very few features were statistically significant in our regression analysis. There was a high Pearson correlation coefficient between tech and \chatbot savviness ($0.6$), the two security and privacy concern questions ($-0.5$, one is negated, the other not), and both savviness and being allowed to use a \chatbot ($0.4$). This may have affected our analysis. There is also the possibility that the predictors were not truly predictive, or that the features affecting these behaviors were not included in our questionnaire.

\subsection{The Threat Surface of CAs Needs to Be Immediately Addressed} 
From a technology perspective, AI security researchers have long demanded to proactively address potential threats~\cite{biggio2018wild}. The potential threat surface identified in our study should therefore be mitigated by vendors as fast as possible \emph{before} threats manifest, to avoid past mistakes and pitfalls regarding AI security evaluations~\cite{carlini2019evaluating,biggio2018wild}. Our study underlines the importance of ongoing efforts against prompt injections~\cite{piet2024jatmo,yi2025benchmarking,yao2024survey} and other vulnerabilities like remote code execution~\cite{chennabasappa2025llamafirewall,beurer2025design}. Ideally, these works consolidate in evaluation or defence guidelines.

For organizations, the descriptive data from our study is an initial starting point to conduct risk assessments for CAs~\cite{ACSFBS24}. Notably, at work we observed more frequent potentially insecure inputs and program access (Sect.~\ref{sec:untrustedInputs}-\ref{sec:securityInputsAndProgram}) compared to leisure contexts. As regular employees engage in ``jailbreaking,’’ models deployed in settings where jailbreak outcomes are security- or privacy-relevant should be secured appropriately~\cite{xie2023defending,yao2024survey} or possibly not be deployed at all. For example, if enterprises employ \chatbots with access to sensitive data (e.g., client, HR, or financial data), technical guardrails are necessary to enforce a need-to-know principle. Moreover, these guardrails need to not only handle large volumes of data, but also large volumes of jailbreak attempts. Equally important are adapted security policies to create awareness among employees not to retrieve or share sensitive information with CAs (Sect.~\ref{sec:privacySharedData}). These have been missing in the past, which encouraged some employees to develop workarounds to avoid disclosing sensitive information~\cite{kimbel2024chatgpt}. 

\rOne{For security practitioners, the main implication is not that every uploaded file, shared data item, or refusal-circumvention attempt results in a successful attack. Rather, behaviors assumed in technical threat models occur frequently enough to be relevant for organizational risk management. CA deployments should therefore be assessed not only in terms of model capabilities, but also in terms of user workflows, tool integrations, data-entry practices, and the availability of guardrails or monitoring mechanisms. Special care should be taken for shadow AI, where threats may emerge although the usage of AI was not known.}

Individuals should be better protected from sharing personal data or malicious content via education, better transparency of vendors, technical or regulative measures. Our results are presumably a conservative estimate for data sharing, as users might share data mostly unknowingly via longer CA conversations that allow the inference of personal data~\cite{piao2021privacy, salem2023sok}. User may have also difficulties to picture privacy risks~\cite{Zufferey2025Popets} and have misconceptions of CAs~\cite{zhang2023s}. Therefore, it is necessary to educate users about the risks of deliberate and unconscious data sharing with CAs. It is also necessary to directly involve the CA vendors to remove the information asymmetry: previous works have shown that natural language interfaces favour sharing of sensitive information, since CAs remind users of interactions with humans~\cite{zhang2023s,kran2025darkbench}. Some CAs reinforce this perception actively by employing \emph{dark patterns}~\cite{kran2025darkbench}. For example, legislation could enforce that the default is that data may not be used and ideally not processed by the provider. Furthermore, providers could be forced to set defaults such that CAs cannot be integrated with programs unless a warning is displayed to the user. While the design of such warnings is a challenge~\cite{bravo2013your}, it may partially alleviate the problem. Another potential approach is to technically restrict input of sensitive data, for example, via prompt sanitization~\cite{Chong2024Casper}.

\subsection{Future Work: Understanding Specific Contexts and Behavioral Predictors} 
Our study provided a first snapshot of selected security- and privacy-related behaviors. In the future, more specific contexts need to be understood, for example, the difference between uploading internal company documents versus external, untrusted ones, or between sharing sensitive company data with corporate CA editions or using private accounts. For use cases in private contexts, analyzing public conversation corpora could complement self-reports and serve as a starting point (e.g., the WildChat corpus~\cite{zhao2024wildchat}), as most CAs are proprietary. Also, what explains and predicts specific behaviors needs to be understood. For example, in the case of information seeking, the behavior may relate to built-in boundaries for provided information or the CA’s poor performance, or be a sign of frustration by end users due to poor usability. The exact nature of these behaviors is yet to be explored and understood, specifically when the \chatbot does not perform according to expectations. Probably not all cases of jailbreaking have to be ``strictly defended’’ using guardrails; in some cases, it may be sufficient to employ less strict constraints, similar to an experimental mode of the CA, after communicating so. In this case, HCI research could support the development of corresponding modes and their communication to the end user.

\section{Conclusion}\label{sec:conclusion}
 With a third of the UK adult population in our sample using \chatbots, security and privacy are increasingly relevant due to the number of users potentially affected. \rTwo{Potential risk-inducing behaviors were reported such as using non-self-created input or} connecting the \chatbot to other programs. Furthermore, a fourth of our sample engages in so-called jailbreaking, although reported motivations are entertainment, understanding limitations, and information searches. The findings highlight the complexity of offering \chatbots as a company, as playful behavior may be hard to discern from real attacks. Finally, while users are aware of privacy risks related to \chatbots, a few still share highly sensitive data. \rTwo{However, many users might share data unknowingly via conversations.} Many \otherChange{users} are also unaware of opt-out mechanisms. Consequently, academic threat models manifest in the wild, and mitigations or guidelines for secure and privacy-friendly usage of \chatbots should be developed. Furthermore, in security and privacy-critical areas, organizations must equip \chatbots with effective AI guardrails to prevent jailbreaking---if they want to use \chatbots at all. \chatbot vendors need to do more to prevent sharing of sensitive data (e.g., passwords). In addition, more transparency is required regarding the use of user data for model training (and ways to opt-out).

\section*{Acknowledgements}
This research was funded by SNSF Grant 207550.

\bibliographystyle{IEEEtran}
\bibliography{lit}

\section{Appendix}\label{app:questions}
\subsection{Screening questions}
\begin{itemize}
\item How often do you use chatbots? By chatbot, or virtual assistants like ChatGPT, Copilot (for example integrated in Bing), Llama, or similar programs, we refer to a software application or web interface designed to mimic human conversation through text or voice interactions. \Chatbots or virtual assistants can solve small tasks like generating images or writing poems. ("Daily", "Weekly", "Monthly", "Once a year")
\item How old are you? ("18-34", "35-54", "55+")
\item What gender do you identify mostly with? ("Male", "Female", "Non-binary / third gender", "Prefer not to say")
\end{itemize}

\subsection{Main questions}
\subsubsection{Demographics and General Questions}
\begin{itemize}
\item When do you use a generative AI \chatbots such as ChatGPT, Copilot, Llama, Copilot or others the most? ("At work", "In my free time", "Both at work and in my free time", "I don't use Chatbots")   
\item What is the highest education level you have completed? (Six-point scale anchored with "No formal qualifications" to "Doctor of Philosophy, PhD, MD or higher")
\item How would you rate your overall level of comfort and proficiency with using technology, such as computers, smartphones, software applications, and the internet? (Five-point scale anchored with "Very Low" and "Very High")
\item How would you rate your overall level of comfort and proficiency with using chatbots? (Five-point scale anchored with "Very Low" and "Very High")
\item When did you first use a chatbot? ("In the last week", "In the last three months", "In the last year", "More than one year ago")
\item Which chatbot(s) do you use? (Mutiple answers; "ChatGPT", "Copilot", "LLama", …, "Other")
\item Please rate your agreement with the following privacy and security statements (Five-point scale anchored "Strongly agree" and "Strongly disagree", Two items: "I am not concerned about threats to my privacy and security today.", "I am uneasy about the current amount of privacy and security I have with my chatbot")
\end{itemize}
The following questions relate to the use of \chatbots at work and/or in leisure time (depending on the participants answer in the first question)
\begin{itemize}
\item Are you allowed to use \chatbots at your workplace? ("No, \chatbots are blocked/forbidden", "Yes, if complying with guidelines", "Yes, without constraints", "I don't know") 
\item How often do you use your \chatbot at work each week on average? Please select the most frequent ("Several times a day", "At least once per day", "Several times per week", "Once per week")
\item Why do you use \chatbots at work? ("To increase productivity", "To enhance creativity", "To search for information", "To receive help for social interactions", "Out of curiosity for a new technology", "For entertainment", "Other")
\end{itemize}
\subsubsection{Specific Questions: Insecure Inputs and Program Access}
\begin{itemize}
\item Do you load any of the following content into \chatbots at work? ("Yes, Images", "Yes, documents (e.g., pdf, word)", "Yes, full websites or links", "Yes, text", "Yes, audio files", "Yes, other", "No")
\item While at work, do you load content into your \chatbot that you did not create yourself (e.g., documents or images from the internet)? ("Yes", "No","I don't know ")
\item Do your give your used \chatbots at work direct access to any of the following software? ("Yes, my email program", ..., "Yes, other", "No nothing").
\end{itemize}
\subsubsection{Specific Questions: Jailbreaking}
\begin{itemize}
\item Has it ever happened to you that the \chatbot intentionally refused an answer or a task? For example, you wanted to know something and the \chatbot told you it is not allowed to answer the question. ("Yes, frequently", "Yes, once", "No, never", "I am not sure")  
\item How did you notice that the \chatbot intentionally refused an answer or a task? ("Chatbot continuously did not provide desired information", "Chatbot's policy/terms of use was shown","Chatbot threw an error","I don't know", "Other")
\item Have you ever tried to make your \chatbot output something it intentionally refused to at first? ("Yes, frequently", "Yes, once", "No, never", "I am not sure")
\item What was your motivation for trying to make your \chatbot output something it intentionallyrefused at first? ("Fun or entertainment", "To obtain desired information", "To explore capabilities and/or explore boundaries", "Other")
\end{itemize}
\subsubsection{Specific Questions: Sensitive Inputs}
\begin{itemize}
\item Do you edit chatbots’ inputs for privacy or security (by, for example, changing names) at work? ("Yes, always when possible", "Yes, sometimes", "Yes, but rarely", "No, never", "I don't use my \chatbot with sensitive data").
\item Which information do you share with your \chatbot how often at work? Please provide for each information shared the frequency as listed. (Six-point scale anchored with "Very often" and "Never" / "Don't know", Twelve items from "Bank account or credit card numbers" to "Customer or client information").
\end{itemize}
The following questions are asked of participants regardless of their work or leisure time usage of chatbots:
\begin{itemize}
\item Who can see the information you shared with your chatbots? ("Only myself", "Employees of the chatbot's parent company", "Other users of the same chatbot", "Potentially everyone on the internet", "I don't know")
\item Does the chatbot's parent company use your inputs to the \chatbot to build a new version of the \chatbot giving better responses? ("Yes", "No", "I don't know")
\item Is it possible to opt out of your chatbot’s parent company using your inputs to the bot to build a new version of the \chatbot giving better responses? ("Yes", "No", "I don't know")
\end{itemize}

\subsection{Detailed regression results}
\rTwo{For program access, the coefficient for allowed was 0.436, with an effect size of 1.53. The cofnidence intervals are 0.232 (lower) and 0.62 (upper). The p-values of all other attributes are larger than 0.23. Model diagnostics are McFadden (0.04), AIC (645.2), and BIC (676). For all program access and senstiive information access regression analysis including the non-significant ones, the model diagnostics where similar and ranged between 0.01 and 0.04 (McFadden), and around 590-762 (AIC/BIC).}

\rTwo{For the regression at work and the second concern question, the coefficient was -0.21, the effect size 0.81, the lower confidence interval -0.327, and the higher confidence interval -0.095. The model diagnostics in this case were 
McFadden (0.035), AIC (1138) and BIC
BIC (1187). For all regression settings, these dignostics ranged within 0.015-0.037 (McFadden) and 1138-1576 (AIC/BIC).}

\end{document}